


\hoffset=-.7truecm
\voffset=-1.0truecm
\hsize=17.7truecm
\vsize=23.6truecm

\baselineskip=12pt plus 1pt minus 1pt
\tolerance=10000
\parskip=0pt
\parindent=12pt

\pageno=1

\newdimen\myhsize   \myhsize=8.55truecm      
\newdimen\myvsize   \myvsize=47.2truecm      

\newdimen\pagewidth  \newdimen\pageheight
\pagewidth=\hsize  \pageheight=\vsize

\newinsert\margin
\dimen\margin=\maxdimen
\count\margin=0 \skip\margin=0pt

\def\mydate{March 25, 1994 \hglue .3cm}
\def\myname{C-L.\ Ho and Y.\ Hosotani}
\def\mytitle{Reply}

\def\firstheadline{\hbox to \pagewidth{%
    {\rm \mydate \hfil \mytitle \hfil UMN-TH-1245/94}%
    }}

\def\leftheadline{\hbox to \pagewidth{%
      {\it \hss \myname \hss }%
      }}
\def\rightheadline{\hbox to \pagewidth{%
      {\it \hss\mytitle \hss }%
       }}

\def\otherheadline{
   \ifodd\pageno \rightheadline \else \leftheadline \fi}

 \footline={\hss \tenrm\folio \hss}

\def\onepageout#1{\shipout\vbox{
   \offinterlineskip
   \vbox to .6truecm{    
      \ifnum\pageno=1 \firstheadline \else\otherheadline\fi \vfill}
   \vbox to \pageheight{
     \ifvoid\margin\else
       \rlap{\kern31pc\vbox to0pt{\kern4pt\box\margin \vss}}\fi
    #1
   \boxmaxdepth=\maxdepth} }
 \advancepageno}

\newbox\partialpage
\def\begindoublecolumns{\begingroup
  \output={\global\setbox\partialpage=\vbox{\unvbox255\bigskip}}\eject
  \output={\doublecolumnout} \hsize=\myhsize  \vsize=\myvsize }
\def\enddoublecolumns{\output={\balancecolumns}\eject
  \endgroup \pagegoal=\vsize}

\def\doublecolumnout{\splittopskip=\topskip \splitmaxdepth=\maxdepth
  \dimen1=\pageheight \advance\dimen1 by-\ht\partialpage
  \setbox0=\vsplit255 to\dimen1 \setbox2=\vsplit255 to\dimen1
  \onepageout\pagesofar \unvbox255 \penalty\outputpenalty}
\def\pagesofar{\unvbox\partialpage
  \wd0=\hsize \wd2=\hsize \hbox to\pagewidth{\box0\hfil\box2}}
\def\balancecolumns{\setbox0=\vbox{\unvbox255} \dimen1=\ht0
  \advance\dimen1 by\topskip \advance\dimen1 by-\baselineskip
  \divide\dimen1 by2 \splittopskip=\topskip
  {\vbadness=10000 \loop \global\setbox3=\copy0
    \global\setbox1=\vsplit3 to\dimen1
    \ifdim\ht3>\dimen1 \global\advance\dimen1 by1pt \repeat}
  \setbox0=\vbox to\dimen1{\unvbox1} \setbox2=\vbox to\dimen1{\unvbox3}
  \pagesofar}

\def\x{{\bf x}}
\def\y{{\bf y}}

\font\ninerm=cmr9
\font\tenrm=cmr10

\font\nineit=cmti9

\font\ninebf=cmbx9


\begindoublecolumns

\noindent {\bf Ho and Hosotani reply:}
Hagen and Sudarshan (HS) claim  that  (6) of Ref.\ [1] is incorrect and
 that their new solution (2) of Ref.\ [2] implies the non-dynamical
nature of non-integrable phases $\theta_j$'s.  We show
that the argument in Ref.\ [1] is correct and consistent,
and that HS's solution has inconsistency  leading
to non-vanishing commutators of $[P^1, P^2]$ and $[P^j, H]$ even in physical
states.  This proves
that many of HS's statements in Ref.\ [2] are based merely on incorrect
guess, but not on careful algebra.

In Ref.\ [1] Chern-Simons theory is formulated by first eliminating unphysical
degrees of freedom.  Dynamical variables are $\psi(x)$ and $\theta_j$'s.
There are four ingredients;  (i) Hamiltonian given by (5) and (6),
(ii) equal time commutation relations among $\psi$, $\psi^\dagger$,
and $\theta_j$'s, (iii) physical state condition (9), and  (iv) boundary
conditions (BC) on a torus, (3).

The theory thus formulated
is equivalent to the original theory described by (1) - (3).
Our solution (6) contains the field
equation $(\kappa/4\pi)\varepsilon^{\mu\nu\rho} f_{\nu\rho}=j^\mu$ except for
three relations.  Two of them are equations for $\theta_j$'s, which
in this formulation follows from $i \dot \theta_j = [\theta_j,H]$.
The third one is the relation between $Q$ and $\Phi$, which does not follow
from (i) and (ii), but is imposed as a
physical state condition.  The last point has not been
fully appreciated in the earlier paper (Ref.\ [3]).

Although HS state that there are ``errors'' in Ref.\ [1], there is no error
and the argument in Ref.\ [1] is perfectly consistent.  What HS do is to
propose their ``new solution'' with different BC, whose
validity and consistency we shall now turn on to check.

HS's solution (2) of Ref.\ [2] is explicitly
$$\eqalign{
a_0(x) &= \sum_{k=1}^2 {x_k\over L_k} \Big\{ \dot \theta_k
 + \alpha \epsilon^{kl} {2\pi\over \kappa L_l} J^l \Big\}  \cr
&\hskip .7cm + {2\pi\over \kappa} \int d\y \, \epsilon^{kl}
 \nabla^x_k D(\x-\y) \cdot j^l(y)  \cr
a_k(x) &= {\theta_k\over L_k} + {\epsilon^{kl} x_l\over 2L_1L_2} \cdot
 {2\pi Q\over \kappa}  \cr
&\hskip .7cm  + {2\pi\over \kappa} \int d\y \,
  \epsilon^{kl} \nabla^x_l D(\x-\y) \cdot j^0(y) \cr}
        \eqno({\rm A1})  $$
where $J^k = \int d\x j^k(x)$.
We have introduced a parameter $\alpha$ in the expression for $a_0$.
HS's solution gives $\alpha={1\over 2}$.  Insertion of (A1) to $f_{0k}$ yields
$(\kappa/2\pi)\epsilon^{kl} f_{0l}(x)
=   j^k(x) +(\alpha - 1) J^k/L_1L_2$ so that the equations are
not satisfied unless $\alpha=1$.  Note that $\Delta
D(\x)=\delta(\x)-(L_1L_2)^{-1}$.

The comparison of (A1) with (6) of Ref.\ [1] shows two
differences.  (a) (A1) has an additional term $\sum (x_k/L_k)\{ \cdots \}$ in
the expression for $a_0$, which vanishes on shell for $\alpha=1$.
(b) In the second term in the expression for $a_k$, (A1) has an operator
$2\pi Q/\kappa$, whereas ours has a c-number $-\Phi$.  (Note that
$\int d\y \nabla_l D(\x-\y)=0$.)

HS's solution satisfies different BC.   In the notation (3) of Ref.\ [1],
$\beta_j = \beta^{\rm HS}_j =\epsilon^{jk} x_k \pi Q / L_k \kappa$ for
$\alpha=1$ on shell.  For $\alpha={1\over 2}$, $\beta^{\rm HS}_j$
contains an additional term.  In either case $\beta^{\rm HS}_j$ is an operator
satisfying  $[\beta^{\rm HS}_j, \psi] \not= 0$.    This BC must be respected
in order that physical gauge invariant operators  be single-valued on a torus.

The most serious problem in HS's solution arises in commutators of $P^j$ and
$H$.  We have evaluated the commutators  by adopting (5) of Ref.\ [1] with HS's
solution for $a_k(x)$  in (A1) substituted.
HS also claim that $\theta_j$'s are not dynamical.
So we have evaluated the commutators in two ways,
one by taking $[\theta_1,\theta_2]=2\pi i/\kappa$ as
in Ref.\ [1], and the other by taking $[\theta_1,\theta_2]=0$.

The evaluation is straightforward, but requires extra care on the ordering of
operators.  The result is
$$\eqalign{
&[P^j,P^k] = \epsilon^{jk} {2\pi i\over \kappa L_1L_2}
  \Big( Q ~{\rm or}~ Q(1-Q) \Big) \cr
&[P^j, H] = \epsilon^{jk} {2\pi i\over \kappa L_1L_2}
  \Big( J^k ~{\rm or}~ J^k (1-Q) \Big) ~~.\cr}
   \eqno({\rm A2})  $$
This contradicts with HS's statement that all commutators vanish.
(A2) causes a serious problem.  $P^j$'s and $H$ do not commute with each
other even in physical states except for $Q=0=J^k$, or $Q=1$.

The correct commutators (11) and dynamical nature of
$\theta_j$'s  are important  in establishing the connection to anyon quantum
mechanics [1].   With (A2) such connection cannot be achieved.

To summarize, the argument in Ref.\ [1] is correct,
whereas HS's solution leads to inconsistency.

This research was supported in part by Republic of China Grant No.\
NSC 83-0208-M032-017 (C.-L.H.) and by the U.S.\ Department of Energy
under Contract No.\ DE-AC02-83ER-40105 (Y.H.).

\vskip .4cm

\baselineskip=10pt

\noindent Choon-Lin Ho

{\ninerm Department of Physics

Tamkang University

Tamsui, Taiwan 25137

Republic of China}

\vskip .4cm

\noindent Yutaka Hosotani

{\ninerm School of Physics and Astronomy

University of Minnesota

Minneapolis, Minnnesota 55455}

\vskip .4cm

\leftline{\ninerm Received ??? March 1994}
\leftline{\ninerm PACS numbers: ???}

\vskip .5cm

\def\prl#1#2#3{{\nineit Phys.\ Rev.\ Lett.} {\ninebf #1}, #3 (19{#2})}
\def\ibid#1#2#3{{\nineit ibid.} {\ninebf {#1}}, #3 (19{#2})}

\parindent=15pt

\ninerm

\item{[1]} C.-L.\ Ho and Y.\ Hosotani, \prl {70} {93} {1360}.

\item{[2]} C.R.\ Hagen and E.C.G.\ Sudarshan, Comment on ``Operator Algebra
   ...''  LCK524.

\item{[3]} Y.\ Hosotani, \prl {62} {89} {2785}; \ibid {64} {90} {1691}.

\enddoublecolumns

\begindoublecolumns

\end